# The Intertwined Rise of Collaboration Scale, Reference Diversity, and Breakthrough Potential in Modern Science: A 40-Year Cross-Disciplinary Study

Sarah J. James[1], Marcus A. Rodriguez[2], and David P. Miller[3]

1 Department of Computer Science, Harvard University, Cambridge, MA 02138, USA.

2 Department of Computer Science, San Jose State University, San Jose, CA 95192, USA.

3 Department of Computer Science and Engineering, University of Toledo, Toledo, OH 43606, USA.

**Abstract**

Over the last four decades, the way knowledge is created in academia has transformed dramatically: research teams have grown larger, scholars draw from ever-wider pools of prior work, and the most influential discoveries increasingly emerge from complex collaborative efforts. Using a massive dataset of over 15 million publications spanning 1970–2010 and covering six major domains (Humanities, Social Sciences, Agricultural Sciences, Medical & Health Sciences, Engineering & Technology, and Natural Sciences), this study tracks how three core features of scientific papers—authorship team size, the breadth and variety of cited sources, and eventual citation impact—have co-evolved over time. We uncover striking differences across disciplines. In every field, papers that build on a broader and more diverse knowledge base consistently attract more citations later on, lending large-scale empirical support to theories that view scientific breakthroughs as outcomes of novel recombination across distant ideas. Bigger teams, on average, generate work with greater ultimate influence, but the gains taper off after a certain scale; very large consortia seldom produce the absolute highest-impact papers. While the Humanities and Social Sciences remain anchored in solo or small-group authorship traditions, the Natural Sciences, Medicine, and Engineering have moved decisively toward big-team "mega-science." These patterns illuminate the underlying production technology of discovery, reveal discipline-specific barriers to collaboration and idea integration, and offer evidence-based guidance for research funding agencies, universities, and policymakers seeking to organize scientific work for maximum breakthrough potential.

## 1 Introduction

The pursuit of scientific knowledge is a complex and challenging endeavor that necessitates collaboration among researchers to answer intricate questions and develop novel insights [1-3]. In recent years, the scientific community has observed a trend towards larger research teams, prompting inquiries about the relationship between team size [4-6], knowledge foundation [1, 7-9], and impact [10-12] in science. This relationship can be conceptualized as a coevolutionary process, in which changes in one variable may impact the others in a dynamic and reciprocal manner. While larger teams have the potential to access more resources and diverse perspectives, facilitating the production of more impactful research, they may also face coordination difficulties that can impede the quality of their work [13-17]. Knowledge foundation also plays a crucial role in determining team impact, with research that builds upon a strong foundation more likely to generate impactful results. Consequently, research teams with a strong foundation in their field may have a higher probability of producing impactful research, regardless of their size [9, 18]. Despite these endeavors, little is known about the dynamic patterns, particularly the interplay relationships of team size, knowledge foundation, and impact in science in the last half-century.

To understand the coevolution of team size, knowledge foundation, and impact in science, this paper analyzes a comprehensive bibliometric dataset, Microsoft Academic Graph [19, 20], and apply statistical and network analysis techniques to identify and analyze the dynamic coevolution patterns of team size, knowledge foundation, and impact of scientific papers over time. By examining this dynamic process, researchers can gain a better understanding of the factors that contribute to successful scientific collaboration and knowledge generation, which can inform strategies to promote impactful research in the future.

This paper aims to investigate the coevolution of team size, knowledge foundation, and impact in scientific research by addressing the following research questions:

(RQ1) How has the growth trend and coevolution of team size, knowledge foundation, and impact of scientific papers evolved over time across different scientific fields?

(RQ2) How do different types of scientific papers (journal articles, conference papers, and top-tier journal articles) exhibit different patterns of coevolution between knowledge foundation and impact?

(RQ3) Is there an optimal team size for generating impactful research, and how does it vary over time?

(RQ4) How do the patterns of knowledge foundation and impact differ across individual science, small team science, large team science, and super-large team science?

By answering these research questions, this study endeavors to investigate the dynamic coevolution of team size, knowledge foundation, and impact of scientific papers across different fields from 1970 to 2010. The study aims to provide insights into the changing landscape of scientific collaboration and knowledge generation, and the underlying factors that contribute to these trends. The research will explore the coevolution of team size, knowledge foundation, and impact over time, identify the optimal team size for generating impactful research, and investigate the distinct trends and patterns in the coevolution of team size, knowledge foundation, and impact across different scientific fields. The ultimate goal of this study is to provide valuable insights into how scientists and policymakers can optimize their strategies for successful collaboration and impactful research. By contributing to the development of more effective policies and practices in the field of scientific research, this research aims to enhance the quality and impact of scientific work and foster greater scientific progress.

## 2 Literature Review

### 2.1 Innovation

As scientific research becomes more specialized and interdisciplinary, it is increasingly difficult for individual researchers to possess the necessary expertise to address complex research questions [21, 22]. Collaborating with researchers from diverse fields and backgrounds can bring together the necessary expertise to address these complex problems [6, 14, 23]. Collaborative research teams have been found to generate more innovative ideas [24], engage in more effective problem-solving, and produce higher quality research outcomes [25]. The growing importance of team science has been recognized by funding agencies and policy-makers. Collaboration among researchers from different disciplines and institutions is often encouraged, as it can lead to more impactful research outcomes and greater scientific advancements [26, 27]. This has led to the formation of large-scale collaborative research initiatives and the establishment of research centers and institutes that bring together researchers from different fields [17, 28].

The shift towards larger research teams has also been driven by advances in technology and the increasing availability of data. With the growing volume of data being generated in scientific research, the need for specialized skills in data analysis and management has become increasingly important [17, 29]. On the one hand, larger and more diverse teams tend to produce more innovative and disruptive research [14, 30, 31]. On the other hand, increasing team size can also lead to coordination challenges and diminishing returns on research output [25]. Larger research teams with members possessing specialized skills in data management and analysis can be more efficient and effective in their research [14, 32]. Additionally, the trend towards larger research teams has been linked to the globalization of science [33]. Collaborative research across different countries and regions has become more prevalent, allowing for the pooling of resources and expertise to address global scientific challenges. This has led to the formation of international research partnerships and collaborations, with researchers from different countries working together to address complex scientific questions [34].

**2.2 Interdisciplinarity**

The knowledge foundation of scientific papers refers to the breadth and depth of the knowledge base used to inform the research. It encompasses the range of literature sources, theories, concepts, and methodologies used to frame the research question and design the study [1]. It's commonly argued that a broad knowledge foundation could enhance the impact and quality of scientific papers [35, 36]. For example, interdisciplinary research often draws on a wide range of literature sources [18], and papers that drew on diverse and novel sources were more likely to become novel and innovative studies and be recognized as significant contributions to the field [8, 9].

Several studies have investigated the relationship between team size and knowledge foundation in scientific research, and the findings have been somewhat mixed [37-68]. Some studies have suggested that larger research teams tend to produce research that is more interdisciplinary and draws upon a broader range of knowledge domains than smaller research teams. This is because larger teams can bring together individuals with diverse expertise and skill sets, which can enhance the overall knowledge foundation of the research. In that case, the level of collaboration have been found to be positively correlated with the knowledge foundation of scientific papers [25]. Collaborating with researchers from different backgrounds can bring together the necessary expertise to address these complex problems and generate impactful research [15].

However, other studies have suggested that larger teams may actually be at a disadvantage when it comes to knowledge foundation, as they may suffer from communication and coordination problems that can impede the flow of knowledge and hinder the integration of different perspectives and approaches. Additionally, larger teams may be more prone to groupthink and other cognitive biases that can limit creativity and innovation [14].

**2.3 Impact**

The field of science of science, has grown substantially in recent years as researchers and policy-makers alike recognize the importance of understanding the factors that contribute to scientific impact [69-85]. One such factor that has received considerable attention is team size. The relationship between team size and impact, as measured by citation count, is a complex and multifaceted one, with a range of theoretical perspectives and empirical findings informing our understanding of this phenomenon [27, 86, 87].

On one hand, there are arguments to suggest that larger research teams are better equipped to tackle complex research questions and produce high-impact findings [88]. Such arguments draw on the idea that larger teams can bring together individuals with diverse expertise, skill sets, and resources, thereby increasing the likelihood that a given research project will be successful. Additionally, larger teams may be able to produce research outputs more efficiently, as they are able to divide tasks among team members and leverage economies of scale [5, 89]. From this perspective, it is expected that larger research teams will have a positive relationship with impact, with the expectation that the more authors involved in a paper, the more citations it is likely to receive.

On the other hand, there are also theoretical arguments and empirical findings that suggest that larger research teams may actually have a negative relationship with impact. One such argument is that larger teams may struggle with issues of coordination and communication, as team members may have different goals, values, and expectations, leading to conflicts and difficulties in integrating research findings [33, 90, 91]. Additionally, larger teams may face challenges in terms of authorship credit, as individual contributions may be less clearly defined or recognized, leading to disputes over authorship order and responsibilities [92]. Empirical findings in this area have been mixed, with some studies finding a positive relationship between team size and impact, while others find no relationship or even a negative relationship.

Despite the conflicting theoretical arguments and empirical findings, it is clear that the relationship between team size and impact is an important one to consider. As the scientific enterprise continues to grow in complexity and interdisciplinarity, it is likely that research teams will continue to increase in size. Understanding how team size influences impact can therefore inform policies and practices that seek to optimize scientific outcomes and ensure that research resources are used effectively [3, 93].

**3 Data**

**3.1 Microsoft Academic Graph**

This study draws upon the Microsoft Academic Graph (MAG) dataset [94], which is widely recognized as the most extensive bibliometric database of scientific research, surpassing both Web of Science (WOS) and Scopus in size [95]. The MAG dataset utilized in this study includes a vast collection of 204,728,474 documents, including journal articles, conference proceedings, and preprints published between 1800 and 2021. Of these documents, 77,427,320 papers have at least one citation, and 68,347,900 papers have cited at least one other paper. The rich and comprehensive dataset provides a valuable source of citation data to examine the dynamic characteristics and coevolution patterns of team size, knowledge foundation, and impact of scientific papers [19].

To ensure sufficient citation windows for analyzing the dynamic coevolution patterns of scientific paper impact [96], this study focuses on papers published in journals and conferences from 1970 to 2009. The MAG dataset provides publication year, journal ISSN, journal names, and conference names, allowing us to limit our scope to papers published in journals and conferences during the designated timeframe. Additionally, we utilize author ID and author order to calculate the number of authors for each paper.

The MAG dataset automatically extracts over 200,000 fields from the abstracts and titles of the papers, enabling us to categorize scientific fields at a high level per paper. We mapped the MAG subjects to six OECD fields: Humanities, Social Sciences, Agricultural Sciences, Medical & Health Sciences, Engineering and Technology, and Natural Sciences.

**3.2 Descriptive analysis**

The primary objective of this study is to investigate the patterns of peer-reviewed research papers to gain insights into broader patterns of science. To achieve this objective, it is necessary to remove most of the comment articles, letters, editorial materials, and other types of documents that indirectly contribute to knowledge production, are usually not peer-reviewed, and have dissimilar team assembly mechanisms [89]. However, our dataset does not provide information on the type of documents, which presents a challenge in identifying and extracting primary research articles from the dataset. To overcome this challenge, we limit the reference numbers of papers to at least 10, based on the common practice that editorial materials and comments often make very few citations. This leaves us with a total of 14,953,855 journal papers and 782,363 conference papers that are considered primary research articles and are included in our analysis. These primary research articles form the basis of our analysis, allowing us to gain insights into the patterns of peer-reviewed research papers in different scientific fields.

Our focus in this study is to untangle the dynamic patterns and coevolution of team size, knowledge foundation, and impact of scientific papers. To achieve this, we use the number of authors in a single paper to denote team size, as it is an intuitive measure of collaboration [4, 6]. Second, we use the number of references cited in the focal paper to denote the knowledge foundation, as the references cited reflect the knowledge base of the focal paper and hence, the level of knowledge foundation [1, 18]. With the super-large capacity of the MAG dataset, we can quantify and compare the level of knowledge foundation for different papers, which allows us to examine how the level of knowledge foundation affects the impact of the paper. Third, we use the 10-year-window citation count [12, 69] to denote the impact of papers, as citation count is a reliable measure of impact and is widely used to represent scientific influence [3, 11].

Table 1 presents the descriptive analysis of team size, number of references, and 10-year citations of journal papers and conference papers, respectively. The mean value of team size for journal papers is 3.78, which is higher than the mean team size of 3.01 for conference papers. The mean number of references for journal papers is 29.78, which is significantly higher than the mean number of references for conference papers, which is 17.84. The mean 10-year citation count for journal papers is 29.06, which is also higher than the mean citation count for conference papers, which is 17.95. These measures allow us to capture the complex interplay between team size, knowledge foundation, and impact of scientific papers over time.

Through our analysis, we aim to provide valuable insights into how the patterns of collaboration, knowledge foundation, and impact have evolved over time and how they are interrelated. These findings can assist scientists and policymakers in optimizing their strategies for successful collaboration and impactful research and contribute to the development of more effective policies and practices in the field of scientific research.

**Table 1. Description of all papers for our sample**

|  | Journal papers | | | Conference papers | | |
|---|---|---|---|---|---|---|
|  | Team Size | #Reference | Citation$_{10}$ | Team Size | #Reference | Citation$_{10}$ |
| Sample Size | 14,953,855 | 14,953,855 | 14,953,855 | 782,363 | 782,363 | 782,363 |
| Mean | 3.78 | 29.78 | 29.06 | 3.01 | 17.84 | 17.95 |
| Std | 5.61 | 27.78 | 76.78 | 2.23 | 10.34 | 60.96 |
| Min | 1 | 10 | 0 | 1 | 10 | 0 |
| 25% | 2 | 15 | 5 | 2 | 12 | 2 |
| Median | 3 | 23 | 13 | 3 | 15 | 6 |

|  | 75% | 5 | 35 | 31 | 4 | 20 | 16 |
|---|---|---|---|---|---|---|---|
|  | Max | 3,176 | 13,267 | 64,593 | 564 | 1,277 | 12,992 |

## 4 Results

### 4.1 Dynamic growth

In this section, we present the results of our analysis to answer RQ1 and RQ2. Our analysis examines the dynamic growth of team size, knowledge foundation, and impact of scientific papers across various publication types, including journal papers, conference papers, and top-journal papers such as Nature, Science, and PNAS. Fig. 1 illustrates the dynamic growth trends of team size, reference numbers, and 10-year citation count of scientific papers over time observed in the past century, providing insights into the changing landscape of scientific collaboration and knowledge generation.

First, our analysis of team size growth over the past half-century (Fig. 1(a-c)) demonstrates a consistent increase in the team size of scientific papers. This trend towards larger research teams can be attributed to the increasing specialization and complexity of research questions and the need for diverse expertise to address these issues [17, 27]. This finding is in line with previous studies that have highlighted the importance of interdisciplinary collaboration and the need for diverse perspectives to address complex tasks [14, 28, 97].

Second, we examine the level of utilization of knowledge foundation in scientific research and how it has evolved over time. The findings presented in Fig. 1 (d-f) demonstrate a steady growth in the utilization of knowledge foundation for journal articles, while the level of utilization has remained stable for conference proceedings. The trend observed for top journal papers, such as Nature, Science, and PNAS, shows some variations. The steady growth in the level of utilization of knowledge foundation for journal papers suggests that the scope and depth of knowledge required to generate impactful research has been expanding over time [98]. The trend towards the utilization of a broader knowledge foundation in journal papers is likely due to the increasing complexity of research questions and the need for more comprehensive and diverse knowledge to address these questions [99], and is also indicative of the increasing importance of collaborative research. In contrast, the level of utilization of knowledge foundation has remained stable for conference papers, which may be attributed to the strict writing specifications and formatting requirements typically imposed by conference organizers.

The findings for top journal papers reveal some interesting variations. Papers published in Nature and PNAS have seen an increase in reference numbers over time, suggesting a greater level of utilization of knowledge foundation. This is likely due to the interdisciplinary nature of research published in these journals and the need for a broader knowledge foundation to address complex research questions [100, 101]. However, the trend observed for papers published in Science is different, as the reference numbers for Science papers have dropped after 1990, indicating a decreasing level of utilization of knowledge foundation. This trend may be due to the journal's strict criteria for publication, which may limit the scope of knowledge that can be included in papers. Additionally, it may be reflective of the journal's emphasis on publishing research with high impact and novelty, which may not require a broad knowledge foundation [8, 18].

Third, we investigated the impact of scientific papers over time, as illustrated in Fig. 1(g-h). The findings revealed fluctuations and rapid growth in the impact of scientific papers, with some notable differences observed for journal papers, conference papers, and top journal papers. The impact of journal papers demonstrated a rapid growth from 1970 to 2005, followed by a slight decrease after 2005. This trend suggests that scientific research published in journals has been increasingly impactful over time, with a growing number of papers generating significant interest and citation in the scientific community. The slight decrease in impact after 2005 may be attributed to a saturation effect, resulting from the increasing number of papers published in journals [3]. Conversely, the impact of conference papers demonstrated fluctuating rising trends, with the highest level observed in 2001. This trend may be attributed to the selective nature of conference publications, where only the best and most impactful research is accepted for publication [102]. Moreover, conference papers tend to focus on cutting-edge research, which may generate a greater level of interest and citation in the scientific community [1]. The impact of top journal papers, such as Nature and Science, demonstrated more rapid growth over time. This may be attributed to the higher visibility and prestige associated with these top-tier journals, as well as the rigorous standards they impose on research quality. Papers published in these journals are typically groundbreaking and innovative, and their impact can lead to significant scientific advancements and breakthroughs [103]. The observed fluctuations in the impact of scientific papers over time may be influenced by a range of factors, including the nature and scope of research questions, the quality of research methodology, and the significance of research findings [2, 104]. Additionally, the impact of scientific papers may be influenced by broader societal and political factors, such as changes in funding priorities and research priorities [105, 106].

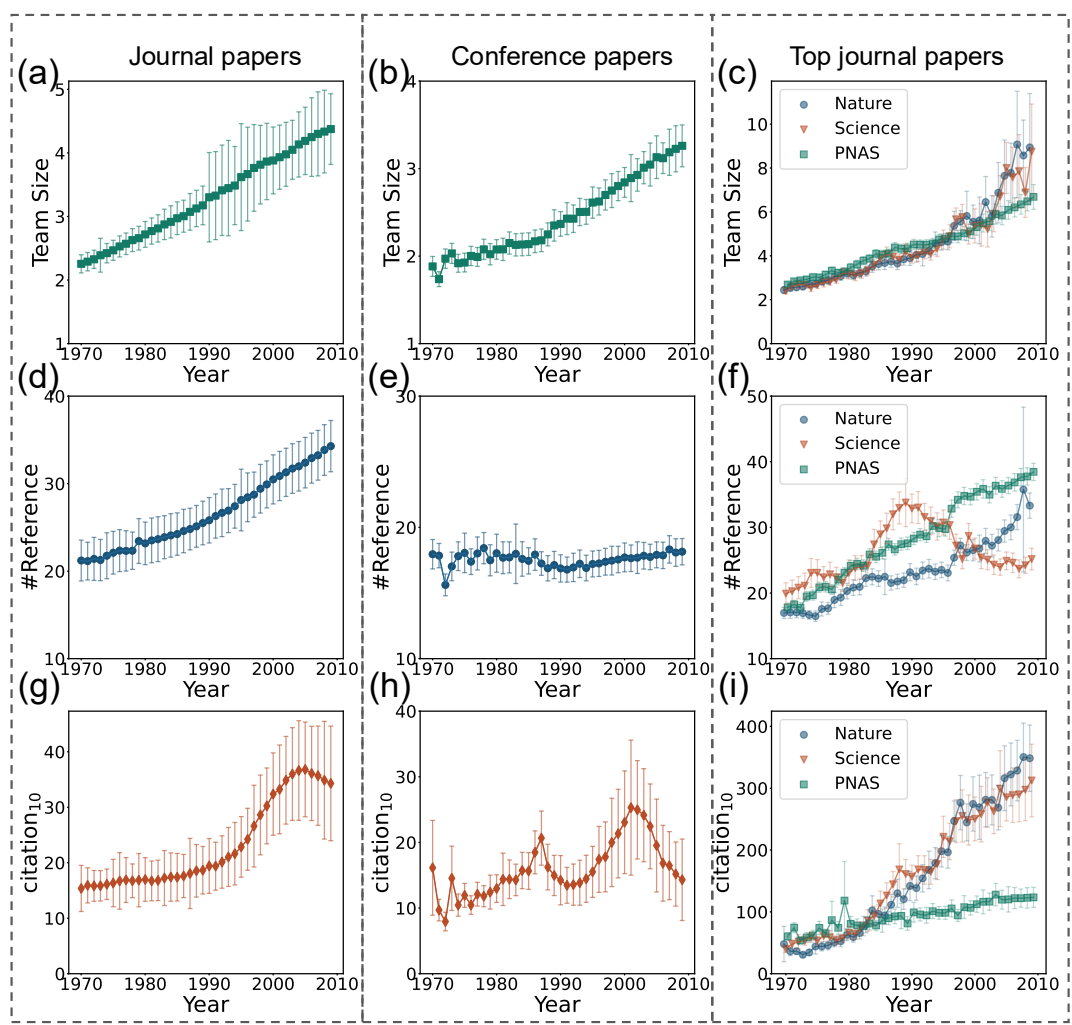

**Fig. 1 Dynamic Evolution of Team Size, Knowledge Foundation, and Impact of Scientific Papers from 1970 to 2010. (a-c)** Increasing Team Sizes of Scientific Papers in Journal Papers, Conference Papers, and Top-Journal Papers. **(d-f)** Widening of Knowledge Foundation for Journal Papers Over Time, While Conference Papers Remain

Stable. **(g-h)** Fluctuation in the Impact of Scientific Papers Over Time for Journal Papers and Conference Papers, with Rapid Growth in Impact for Nature and Science. Error bars represent standard deviation.

**4.2 Knowledge foundation vs impact**

In this section, we examine the coevolution of knowledge foundation and impact in scientific papers, which is a complex and dynamic process influenced by various factors such as research methodology, funding priorities, and societal and political factors. To address RQ1 and RQ2, we further analyze the dynamic coevolution of knowledge foundation and impact over time. The findings, presented in Fig. 2, provide important insights into how the coevolution of knowledge foundation and impact has evolved over time across different publication types.

Overall, our analysis reveals a strong positive correlation between the knowledge foundation of scientific papers and their impact, as observed across different types of papers. The results indicate that a broader knowledge foundation corresponds to higher impact in scientific papers.

Specifically, for journal papers, the impact gap between papers with the most broad-knowledge foundation (with more than 100 references) and those with the narrowest-knowledge foundation (with 10-15 references) has become larger over the past half-century. While the impact of narrow-knowledge foundation papers has shown little growth from 1970 to 2010, the impact of broad-knowledge foundation papers has grown rapidly, from about 40 to about 120 (10-year citations). This trend suggests that a broader knowledge foundation is almost essential for generating impactful research, especially in modern science. Similarly, our findings show that for top-tier journal papers, the impact of broad-knowledge foundation papers contributes much more to impact growth than narrow-knowledge foundation papers. This finding highlights the importance of a comprehensive and diverse knowledge foundation for generating impactful research in top-tier journals, which are known for their strict standards for research quality and innovation. These results provide important insights into how scientists can optimize their strategies for impactful research and how policymakers can develop effective policies and practices to foster scientific progress.

However, the dynamic patterns for conference papers are slightly different. In the period of 1970s, the most broad-knowledge foundation papers (with more than 100 references) have already showed a dominant advantage in impact over lower ones. This suggests that the selection criteria for conference papers may have been different during that period, with a greater emphasis on high-quality collaborative research.

The positive correlation between knowledge foundation and impact can be attributed to the increasing complexity and interdisciplinary nature of scientific research. As research questions become more complex and multidisciplinary, it becomes increasingly challenging for individual researchers to possess the necessary expertise to address these questions. Collaboration among researchers from different backgrounds can bring together the necessary expertise to address these complex problems and generate impactful research. Furthermore, the growing significance of team science has been recognized by funding agencies and policymakers. Collaboration among researchers from different disciplines and institutions is often encouraged, as it can lead to more impactful research outcomes and greater scientific advancements.

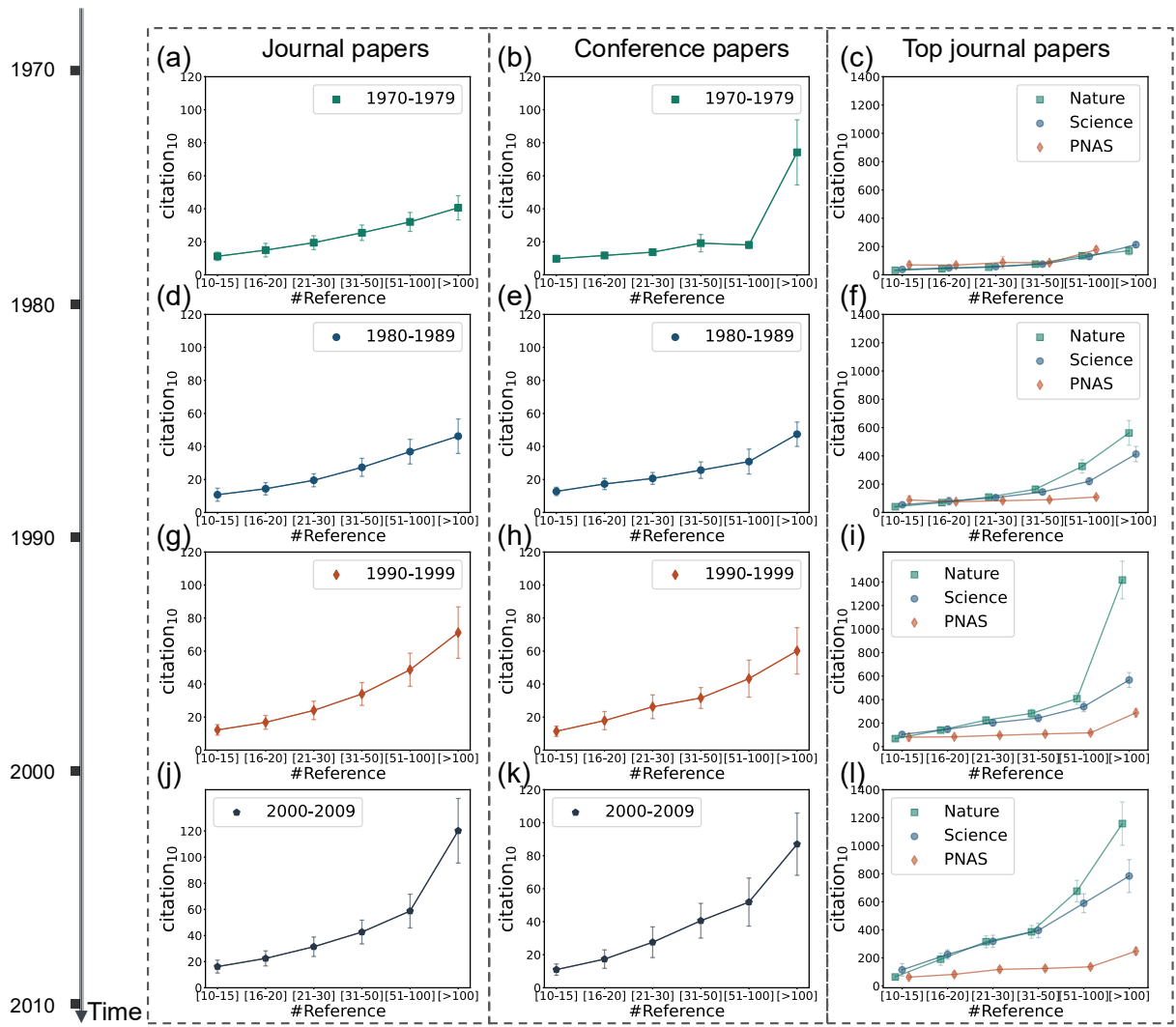

**Fig. 2 Coevolution of Knowledge Foundation and Impact over Time**. **(a, d, g, j)** Binned Reference Numbers and 10-Year Citations for Journal Papers in **(a)** 1970-1989, **(d)** 1980-1989, **(g)** 1990-1999, and **(j)** 2000-2009. **(b, e, h, k)** Binned Reference Numbers and 10-Year Citations for Conference Papers in **(b)** 1970-1989, **(e)** 1980-1989, **(h)** 1990-1999, and **(k)** 2000-2009. **(c, f, i, l)** Binned Reference Numbers and 10-Year Citations for Top Journal Papers (Nature, Science, PNAS) in **(c)** 1970-1989, **(f)** 1980-1989, **(i)** 1990-1999, and **(l)** 2000-2009.

**4.3 Team vs knowledge foundation and impact**

To address RQ3, we delve deeper into the coevolution of knowledge foundation and impact in relation to team size for journal papers. Fig. 3 presents the dynamic coevolution of knowledge foundation and impact versus team size over time, with panels depicting binned team sizes and reference numbers or 10-year citations for different time periods, ranging from 1970 to 2009. Our analysis yields several key findings.

First, the results reveal a changing pattern in the utilization of knowledge foundation by small and large teams over time. In the 1970s, small teams tended to utilize a broader knowledge foundation, while large teams utilized a narrower knowledge foundation. However, in the 1980s and 1990s, large teams experienced a rapid growth in the utilization of knowledge foundation, while small teams showed only slight growth. This trend indicates that large teams were better able to capitalize on the increasing availability of specialized knowledge and expertise to generate impactful research outcomes, while small teams may have faced more challenges in accessing and integrating specialized knowledge into their work. In the 2000s, large teams tended to utilize a broader knowledge foundation, while small teams utilized a narrower knowledge foundation. This suggests that as research questions become increasingly complex, large teams are better equipped to incorporate diverse perspectives and expertise to generate impactful research outcomes, while small teams may struggle to keep up with the growing scope and depth of knowledge required to address these questions effectively [107]. The observed changes in the pattern of knowledge utilization by small and large teams over time are likely due to the evolving nature of scientific research and the changing landscape of scientific communication and publishing [25].

Second, the analysis shows that large team sciences tend to receive higher impact, but this pattern also changes over time. Between 1970 and 1990, large team sciences only had a slightly higher impact than small team sciences, indicating that during this period, small teams were able to compete with large teams in terms of generating impactful research outcomes, despite their smaller size. However, after 1990, large team sciences received much higher impact, pointing to a growing advantage of large teams in generating impactful research outcomes over small teams. The increasing dominance of impact for large team science in recent years corresponds to their high ability to develop knowledge generation strategies that adapt to the changing landscape of scientific research. Moreover, large teams are able to leverage their collective reputations to gain better recognition within scientific communities, which can lead to further opportunities for collaboration and funding [27, 108, 109]. It is important to recognize that this does not necessarily indicate that small teams cannot achieve impactful research outcomes, but rather that large team sciences have gained advantages in generating impactful research outcomes in contemporary scientific research.

Third, to address RQ4, our analysis demonstrates a significant shift in the coevolution of knowledge foundation and impact from individual to large team science over time. Specifically, individual scientific papers utilized a broader knowledge foundation before 2000, yet received the lowest impact. However, after 2000, large team science tended to utilize a broader range of knowledge foundation than individual science, while individual science continued to receive the lowest impact. Despite the significant growth in the utilization of knowledge foundation in individual science from 1970 to 2010, the impact of individual science remained relatively stable. This trend highlights the challenges faced by individual scientists in generating impactful research outcomes due to limited resources and expertise.

Furthermore, our analysis focuses on super-large team science (with more than 100 authors for a single paper). The results demonstrate a sharp growth in the utilization of knowledge foundation and impact for super-large team scientific papers over time. However, the most impactful science is not super-large team science, but large team science (with team sizes of [31-50] or [51-100]). This finding suggests that an optimal team size for impactful research may exist, and that super-large teams may not always be necessary for generating the most impactful research. The observation that the most impactful science is generated by large teams with team sizes of [31-50] or [51-100] highlights the importance of finding the right balance between team size and expertise in scientific research. Larger teams may be better equipped to incorporate diverse expertise, but may also face challenges in coordinating efforts and communicating effectively. Smaller teams may have a more focused and efficient workflow, but may lack the resources and expertise necessary to tackle complex research questions effectively [14]. These findings have important implications for how scientific research is conducted, managed, and supported, and may inform strategies for optimizing team science for impactful research outcomes.

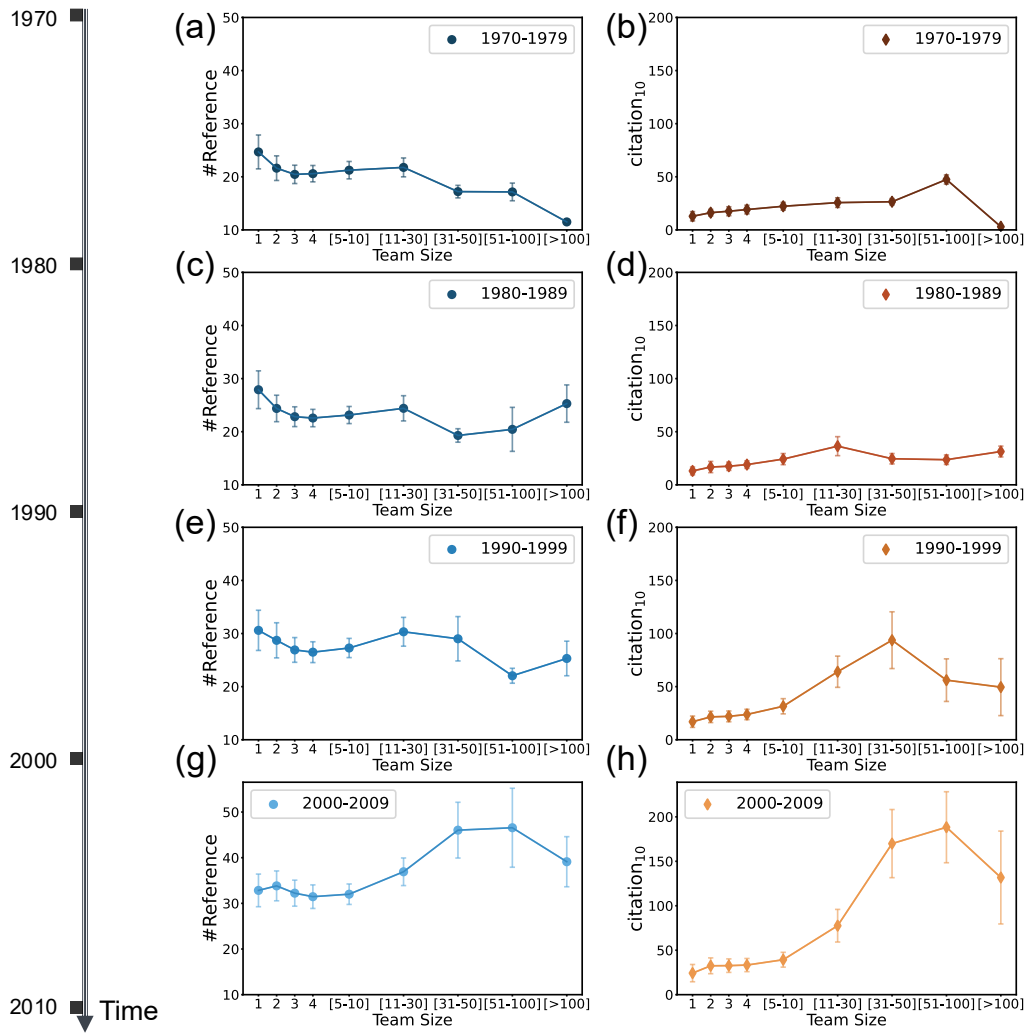

**Fig. 3 Coevolution of Knowledge Foundation and Impact vs Team Size (for Journal Papers) over Time. (a, c, e, g)** Binned Team Sizes and Reference Numbers for Journal Papers in **(a)** 1970-1989, **(c)** 1980-1989, **(e)** 1990-1999, and **(g)** 2000-2009. **(b, d, f, h)** Binned Team Sizes and 10-Year Citations for Conference Papers in **(b)** 1970-1989, **(d)** 1980-1989, **(f)** 1990-1999, and **(h)** 2000-2009.

**4.4 Evolution across field**

In this section, we investigate the coevolution of team size, knowledge foundation, and impact in scientific papers across different fields to answer RQ1. Fig. 4 provides a comprehensive depiction of the dynamic growth of team size, reference numbers, and 10-year citation counts in various scientific fields, which are normalized to the average value of all published papers in each decade. The results show significant variations in the coevolution of these dimensions across different fields, providing valuable insights into the changing landscape of scientific research.

Our analysis demonstrates that Humanities research exhibits a lower tendency towards teamwork and receives the lowest 10-year citation counts when compared to other fields. This finding suggests that Humanities research may prioritize individual scholarship over collaborative work, resulting in lower impact compared to other fields. Nevertheless, the analysis shows that Humanities papers demonstrate a broad range of knowledge foundation, indicating a breadth of scholarship and multidisciplinary approaches in Humanities research. Moreover, the structure of the three dimensions remains stable over time, indicating that the strategies for effective collaboration and knowledge generation may be different in Humanities research. On the other hand, Social

Sciences exhibit a declining trend in collaboration over time but show an increasing utilization of a broad range of knowledge foundation and higher impact from 1970 to 2010. This finding suggests that Social Sciences research may prioritize interdisciplinary approaches over large-scale collaboration, leading to higher impact relative to other fields. However, the decreasing trend in collaboration in Social Sciences may reflect the challenges of coordinating large-scale interdisciplinary research projects in this field [110].

In contrast, Agricultural Sciences demonstrate a growing utilization of knowledge foundation and growing impact over time, with an average level of collaboration compared to all fields. This trend suggests that Agricultural Sciences may have the trend of leveraging multiple approaches and resources to address complex research questions in recent years. Medical & Health Sciences exhibit the highest level of team size, knowledge foundation, and impact, indicating that this field may be the most mature discipline from 1970s. However, their relative utilization of knowledge and impact is gradually decreasing, indicating that Medical & Health Sciences may be facing challenges in maintaining the same level of impact as other field matures, as research questions become more complex and specialized. Engineering Technology demonstrates a high and stable level of collaboration from 1970 to 2010, but does not utilize a broad range of knowledge foundation, indicating a more specialized focus in this field. Notably, their relative impact is increasing rapidly from 1990s to 2000s, suggesting that Engineering Technology may be addressing critical and impactful research questions, and gain broader recognition. Finally, Natural Sciences exhibit a balanced and stable trend with a level of team size, knowledge foundation, and impact that is almost the same as the average of all fields and does not vary much over time. This finding suggests that Natural Sciences may be employing effective strategies for collaboration and knowledge generation, leading to consistent and stable impact over time.

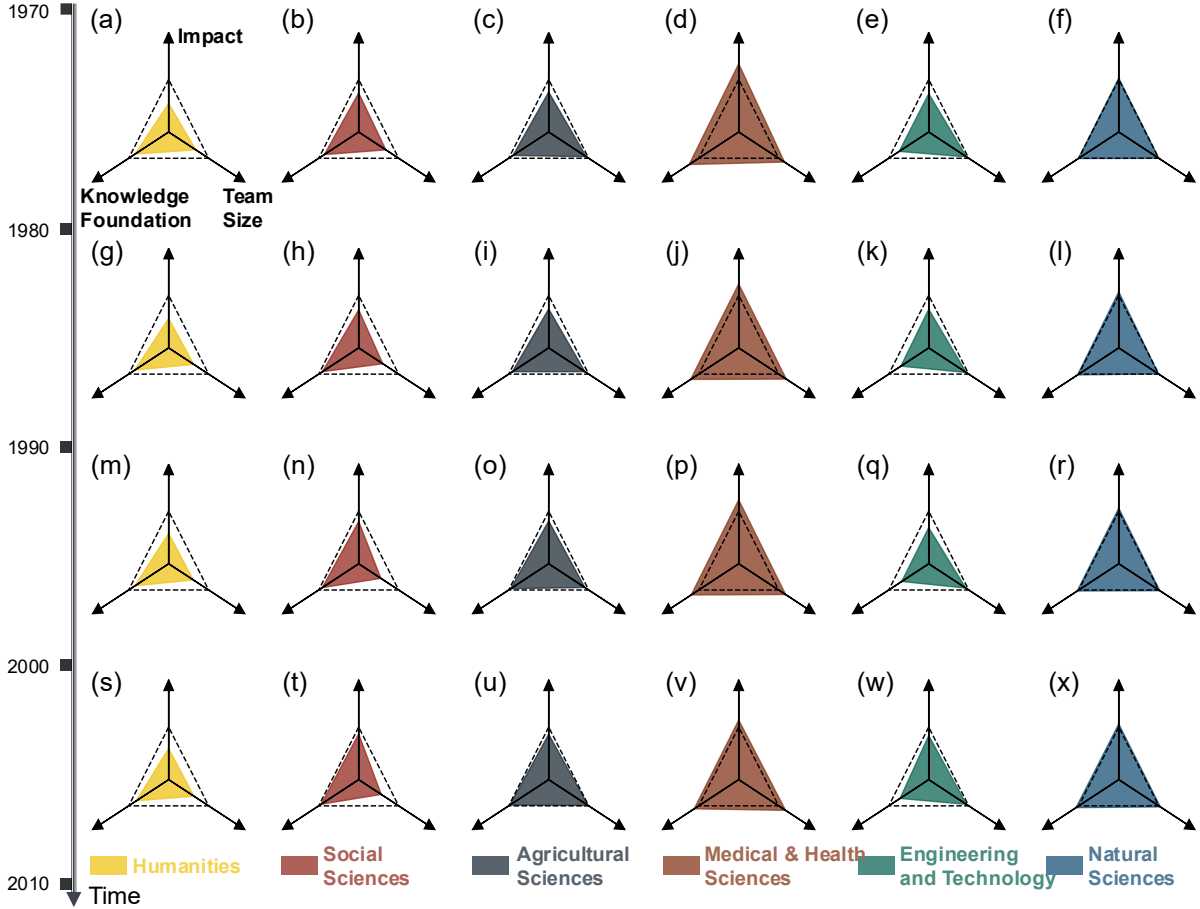

**Fig. 4 Normalized Coevolution of Team Size, Knowledge Foundation, and Impact of Scientific Papers in Different Fields from 1970 to 2010.** The Dashed Triangle Lines Represent the Average Value in Each Decade, While the Filled Area Corresponds to the Relative Value of Each Field for Team Size, Knowledge Foundation, and Impact. We Analyze 6 Fields in MAG: **(a, g, m, s)** Humanities, **(b, h, n, t)** Social Sciences, **(c, i, o, u)** Agricultural Sciences, **(d, j, p, v)** Medical & Health Sciences, **(e, k, q, w)** Engineering and Technology, and **(f, l, r, x)** Natural Sciences.

## 6 Discussion

### 6.1 Theoretical Implications

The theoretical implications of this paper are significant for advancing our understanding of the complex interplay between collaboration, knowledge generation, and scientific impact. This paper sheds light on the evolving trends and patterns of scientific collaboration, highlighting the importance of understanding the dynamics of team size and knowledge foundation in generating impactful research. First, this paper contributes to the literature on collaboration in science by demonstrating the increasing importance of teamwork in generating impactful research [87, 97]. The results indicate that large teams tend to receive higher impact, particularly after 1990, and that an optimal team size for impactful research may exist, while super-large team science may not always be necessary for generating the most impactful research.

Second, this paper contributes to the literature on knowledge generation in science by highlighting the importance of a broad knowledge foundation for generating impactful research [18, 111]. The results indicate that a broader knowledge foundation is almost essential for generating impactful research in recent years, providing evidence that broad knowledge sources could promote impactful research.

Third, this paper contributes to the literature on scientific impact by demonstrating the complex interplay between team size, knowledge foundation, and impact [1, 14, 25]. The results reveal that a broader knowledge foundation of scientific papers corresponds to higher impact, and that large team sciences tend to receive higher impact. However, the pattern also changes over time and across fields, highlighting the need for dynamic approaches to understanding the relationship between team size, knowledge foundation, and impact.

### 6.2 Practical Implications

This study has several practical implications for scientists, policymakers, and funding agencies. The findings can help them optimize their strategies for successful collaboration and impactful research and contribute to the development of more effective policies and practices in the field of scientific research. First, this study highlights the importance of a broad knowledge foundation for generating impactful research, suggesting that scientists and research teams should aim to incorporate a diverse range of expertise and knowledge in their research to achieve high impact. Policymakers and funding agencies can encourage interdisciplinary collaboration and support research teams with diverse expertise to promote impactful research.

Second, this study provides insights into the optimal team size for generating impactful research. The results indicate that large team science tends to receive higher impact than small team or individual science, and an optimal team size for impactful research may exist, yet super-large teams may not always be necessary for generating the most impactful research. Policymakers and funding agencies can consider this finding when designing research funding programs and encourage team science with a moderate size to balance collaboration and efficiency.

Third, this study provides insights into the distinct trends and patterns in the coevolution of team size, knowledge foundation, and impact of scientific papers across different fields. Policymakers and funding agencies can consider these unique dynamics when designing research funding programs and policies for specific scientific fields. In fields where collaboration is rare, such as Humanities and Social Sciences, policymakers and funding agencies can encourage interdisciplinary collaboration to promote impactful research.

**6.3 Limitation and Outlook**

Despite the valuable insights provided by this study, some limitations should be acknowledged. First, this study only focused on scientific papers and did not consider other types of research output, such as patents, books, or technical reports. Future studies could investigate the coevolution of team size, knowledge foundation, and impact in these other types of research output to gain a more comprehensive understanding of the dynamics of research collaboration and knowledge generation. Second, this study utilized a quantitative approach and did not delve into the qualitative aspects of research collaboration and knowledge generation. Future studies could use a mixed-methods approach to explore the factors that contribute to successful collaboration and impactful research. Finally, this study only examined the coevolution of team size, knowledge foundation, and impact at a macro-level. Future studies could investigate these dynamics at a micro-level, such as examining the coevolution of these factors within specific research teams or projects.

In terms of future outlook, this study provides a valuable foundation for further research in the field of scientific collaboration and knowledge generation. The findings of this study can inform the development of policies and strategies to promote effective collaboration and impactful research, including the development of interdisciplinary research teams and the fostering of a diverse and inclusive research environment. Moreover, this study highlights the importance of considering the complex and dynamic interplay between team size, knowledge foundation, and impact in scientific research. As the scientific landscape continues to evolve, understanding these dynamics will be crucial for the successful generation of impactful research.

**Conflict of Interest**

We declare no competing interest relevant to this article.